# From Software Architecture Structure and Behavior Modeling to Implementations of Cyber-Physical Systems

Jan Oliver Ringert[*] and Bernhard Rumpe and Andreas Wortmann

Software Engineering
RWTH Aachen University
Ahornstrasse 55
52074 Aachen, Germany
http://www.se-rwth.de/

**Abstract:** Software development for Cyber-Physical Systems (CPS) is a sophisticated activity as these systems are inherently complex. The engineering of CPS requires composition and interaction of diverse distributed software modules. Describing both, a system's architecture and behavior in integrated models, yields many advantages to cope with this complexity: the models are platform independent, can be decomposed to be developed independently by experts of the respective fields, are highly reusable and may be subjected to formal analysis.

In this paper, we introduce a code generation framework for the MontiArcAutomaton modeling language. CPS are modeled as Component & Connector architectures with embedded I/O$^\omega$ automata. During development, these models can be analyzed using formal methods, graphically edited, and deployed to various platforms. For this, we present four code generators based on the MontiCore code generation framework, that implement the transformation from MontiArcAutomaton models to Mona (formal analysis), EMF Ecore (graphical editing), and Java and Python (deployment). Based on these prototypes, we discuss their commonalities and differences as well as language and application specific challenges focusing on code generator development.

## 1 Introduction

Cyber-Physical Systems (CPS) [Lee06] are distributed interactive systems which combine computational and physical processes. Typical CPS are found in the manufacturing, automotive, smart energy, avionics, and distributed robotics domains. Software development for CPS is a sophisticated endeavor which yields many challenges. The systems are logically and physically distributed, need to perform on diverse platforms, fulfill certain run-time properties, and deal with communication issues.

Component-based software engineering has been applied to tackle these complexities by breaking systems down to platform dependent components [BBC[+]07, NFBL10], thus requiring domain experts to be expert software developers, too. The "accidental complexi-ties" [FR07] arising from this gap between problem domain and implementation domain can be reduced using modeling techniques [SSL11, BR12].

---

[*]J.O. Ringert is supported by the DFG GK/1298 AlgoSyn.

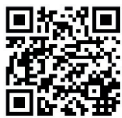

We propose a model-driven approach to engineering of CPS where the systems are modeled as Component & Connector (C&C) architectures using automata to describe the components behavior. These models can be refined from specifications to implementations (supported by formal analysis techniques) and translated into various platform specific implementations. Our approach combines concepts from architecture description languages (ADLs) for modeling the structure of software architectures [TMD09] and behavior description languages [Har87, GBWK09].

In this paper, we introduce a code generation framework for the modeling language MontiArcAutomaton [RRW12]. MontiArcAutomaton extends the ADL MontiArc [HRR12] with behavior description by embedded I/O$^\omega$ automata [Rum96, Mon12]. This language allows distributed and target platform independent modeling of both the structural architecture of the system as well as its behavior. The languages MontiArc and MontiArcAutomaton are developed using the MontiCore framework [KRV10]. We claim using the MontiArcAutomaton modeling language for the development of CPS yields several advantages:

- The use of a C&C architecture description language makes communication and dependencies explicit in the models.
- MontiArcAutomaton allows behavior underspecification in two forms: (1) incompleteness of triggers to only regulate the reaction to inputs of interest and (2) non-deterministically overlapping triggers to restrict possible behavior as desired. These powerful specification mechanisms are supported by automatic verification and refinement checking as presented in [Kir11].
- The logical decomposition of MontiArcAutomaton components allows independent, incremental and bottom-up modeling, and analysis by different domain experts.

In various robotics projects we have developed MontiArcAutomaton (code) generators for EMF Ecore[1] for graphical editing within Eclipse, Mona [EKM98] theories for verification and refinement checking of models and requirements [RRW12] during the development process, and Java and Python code generation to deploy the modeled systems to robots running the educational LeJOS[2] and the industrial ROS[3] platforms. These generators implement the template-based code generation approach of MontiCore [Sch12], which facilitates development of new target language code generators by allowing to reuse great parts of existing code generators.

We illustrate the benefits of model-based development of CPS with MontiArcAutomaton by describing a system of connected robots providing a collaborative mapping service in Section 2. Afterwards, we introduce the MontiCore framework in Section 3 and we describe the MontiArcAutomaton modeling language in Section 4. Subsequently, we introduce and discuss the different code generators in Section 5. We review related work in Section 6 and conclude this contribution in Section 7.

---

[1] The Eclipse Modeling Framework Project: http://www.eclipse.org/modeling/emf/
[2] Java for LEGO Mindstorms http://lejos.sourceforge.net/
[3] Robot Operating System http://www.ros.org/

## 2  Example Software Architecture and Behavior Implementation

An engineer is developing a system of distributed mobile robots to map an a priori unknown office floor. To lower production cost by avoiding expensive sensors, the robots have to estimate their position based on their starting point and performed movement commands. This naive odometric approach to simultaneous localization and mapping (SLAM) [TL08] is prone to produce inaccurate maps, as the difference between estimated position and real position increases over time. The engineer counters this problem by implementing a simultaneous and cooperative discovery by multiple robots communicating via Bluetooth.

The engineer developing the robots has defined a system architecture based on the sensors and actuators available. Figure 1 illustrates the architecture of a single SLAM robot consisting of a front mounted `TouchSensor` component to detect obstacles, a `BumpControl` component implementing the robot controller, a `MapBuilder` component constructing the map from commands passed to the motors and feedback from other SLAM robots received via the `Bluetooth` component.

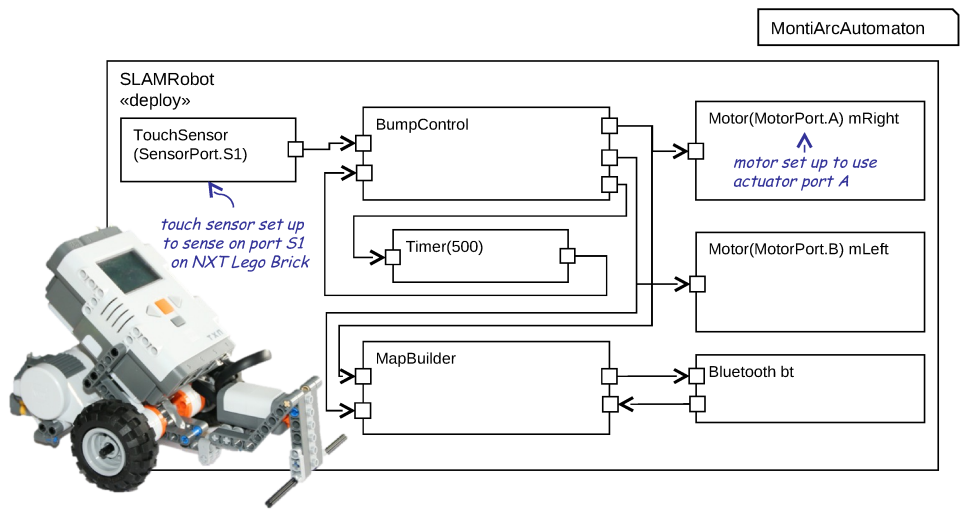

Figure 1: The SLAM robot architecture `SLAMRobot` with a `TouchSensor`, a `Timer`, two motors, the controller `BumpControl`, the component `MapBuilder`, and the Bluetooth communication.

The robot will drive around straight forward until it discovers a new boundary of the map to be explored — by bumping into it. It will then back off and continue the exploration continuously monitoring its own position and communicating with the other robots. For the rest of the example, we focus on the component `BumpControl` that handles the bumping into map boundaries and the subsequent driving maneuvers.

After developing the system architecture, the engineer starts designing the implementation by creating an initial version that defines basic behavior constraints for the `BumpControl` component (e.g., do not drive forward when the bumper is pressed). She later refines it to the implementation shown in Figure 2. According to the implementation the `BumpControl` component starts in state `idle` with both motors stopped. Once the bumper is pressed by the user to activate the robot, it sends a `FORWARD` command to both motors. When the SLAM robot runs into an object, it backs up, turns around and proceeds forward. The backing and turning times are determined by an external timer that is set via a message on port `tc` and responds with an alert via port `ts`.

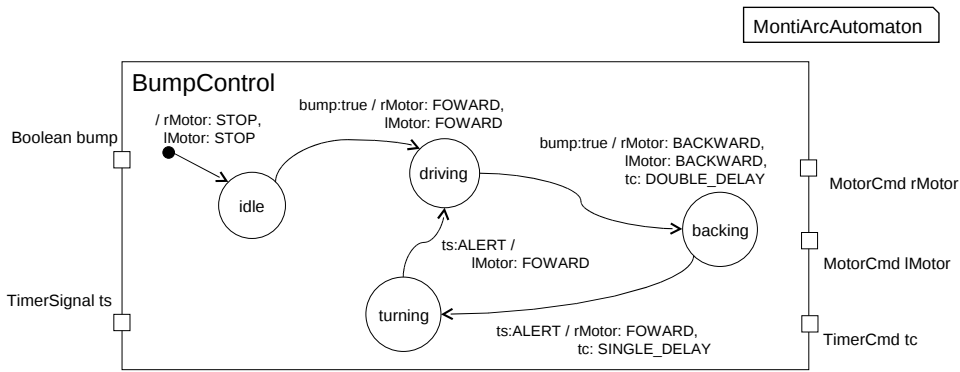

Figure 2: The MontiArcAutomaton component `BumpControl`.

Both the initial specification and the refined implementation are translated to Mona to check refinement, i.e., to check that the implementation does not violate its specification. An excerpt from the translation of the component `BumpControl` is shown in Listing 3.

```
pred bumperbot_BumpControl( var2 bump_true, var2 bump_false,
  # ..., more port values here
  var2 allTime )
  # states of automaton
= ex2 idle, driving, backing, turning:
  # ... constraints: one state at a time
  # initial states and their outputs:
  ( (0 in idle & 0 in rMotor_STOP & 0 in lMotor_STOP ) &
  all1 t: t+1 in allTime => (
    (t in idle & t in bump_true &
     t+1 in driving & t+1 in rMotor_FORWARD &
                      t+1 in lMotor_FORWARD ) |
    # ... more transitions here
  ));
```

Listing 3: An excerpt from the Mona predicate generated from the automaton inside `BumpControl`.

To analyze the model a MontiArcAutomaton code generator translates each component to a predicate over streams of messages. The parameters of the predicate are the possible values on input and output streams of the component over time (see Listing 3, l. 1 for the stream on input port `bump`) and a synchronized time variable `allTime` (l. 3). The encoding of message streams into WS1S logic (Weak Second-order monadic logic of 1 Successor) is inspired by [Sch09]. It uses one second order variable for each possible value on a stream, e.g., `bump_true` and `bump_false` (l. 1). The predicate holds iff the output streams are valid responses to the input streams.

Values on streams and active states are represented by sets of natural numbers, e.g., sets `idle` and `driving` (l. 5). A number $t \in \mathbb{N}$ is in a set iff the corresponding state (or value) is chosen at time $t$. Thus the initial state `idle` is translated to `0 in idle` (l. 8). The transition system is defined analogously for the source state and input at time `t` and the target state and output at time `t+1` (ll. 10-12). For more information on our translation to Mona see [Kir11].

After making sure that the implementation refines its specification the engineer is confident about her work and generates Java code that she compiles and deploys to her robot. Parts of the Java code generated from component `BumpControl` are shown in Listing 4. The displayed excerpt of the Java class `BumpControl` shows how the automaton's initial state (l. 3) and the initial values of the motors (ll. 4, 5) are set based on the initial output as shown in Figure 2. The excerpt of the `compute()` method shows how the first transition from state `idle` to state `going` is translated to Java.

```Java
public class BumpControl implements Component {
  public void init() {
    this.state = State.idle;
    this.rMotor.setCurrentValue(MotorCmd.STOP);
    this.lMotor.setCurrentValue(MotorCmd.STOP);
  }
  public void compute() {
    if (this.state.equals(State.idle)
          && (this.bump.getCurrentValue() != null
          && this.bump.getCurrentValue() == true) ) {
      this.rightMotor.setNextValue(MotorCmd.FORWARD);
      this.leftMotor.setNextValue(MotorCmd.FORWARD);
      this.state = State.driving;
    }
    // ... more transitions here
  }
```

Listing 4: An excerpt from the Java code generated from the automaton of `BumpControl`.

## 3  MonstiCore Language Framework

We have developed the ADL MontiArc [HRR12] and the corresponding framework using the language workbench MontiCore [KRV10]. MontiCore facilitates the development of domain specific modeling languages by providing a grammar for language definition and tools for parser generation, symbol table management, code generation and a context conditions framework. MontiCore languages like MontiArcAutomaton, are defined by context-free grammars. To check properties not expressible in context-free grammars, e.g., whether a variable is defined twice, MontiCore provides a compositional Java-based context condition framework [Vö11]. As these context conditions often require information from other models (e.g., to determine whether an assignment violates a type constraint), MontiCore also contains a compositional symbol table framework [Vö11], to facilitate development of complex context conditions.

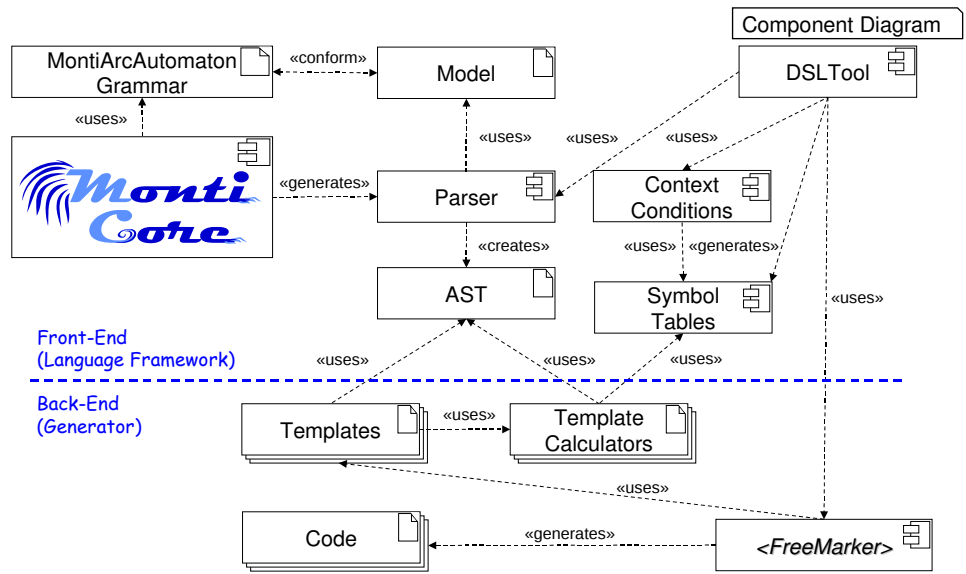

Figure 5: MontiCore uses the grammar to generate a parser for MontiArcAutomaton models which creates the AST (see [Sch12]). The `DSLTool` uses the parser to read models, which are validated by the context condition framework using the symbol tables provided by the `DSLTool`. The `DSLTool` further may use FreeMarker templates and template calculators to generate code from the models based on the AST.

The component diagram in Figure 5 illustrates the components of the MontiCore DSL framework: the compositional approach of MontiCore facilitated development of MontiArcAutomaton by generating a parser for the MontiArcAutomaton DSL and providing frameworks for context conditions, symbol table generation and code generation. The code generation framework of MontiCore [Sch12] utilizes the Java-based template engine FreeMarker[4] to construct new code generators. Using this framework, a new code generator usually only requires a few new calculators and templates. MontiCore further facilitates language and generator development by means of language inheritance and composition mechanisms (like the embedding of I/O$^\omega$ automata into MontiArc). We discuss these, their influence on the code generation, and the reuse of calculators and templates in Section 5.

MontiCore also provides a framework for the generation of text editor Eclipse plugins.

---

[4]Freemarker Template Engine: http://freemarker.org/

This framework offers a DSL which allows to define editor models and a set of workflows which editor developers have to implement for their languages. Using this and the Ecore generation of MontiArcAutomaton, we have developed a both textual and graphical editor for MontiArcAutomaton models as displayed in Figure 6.

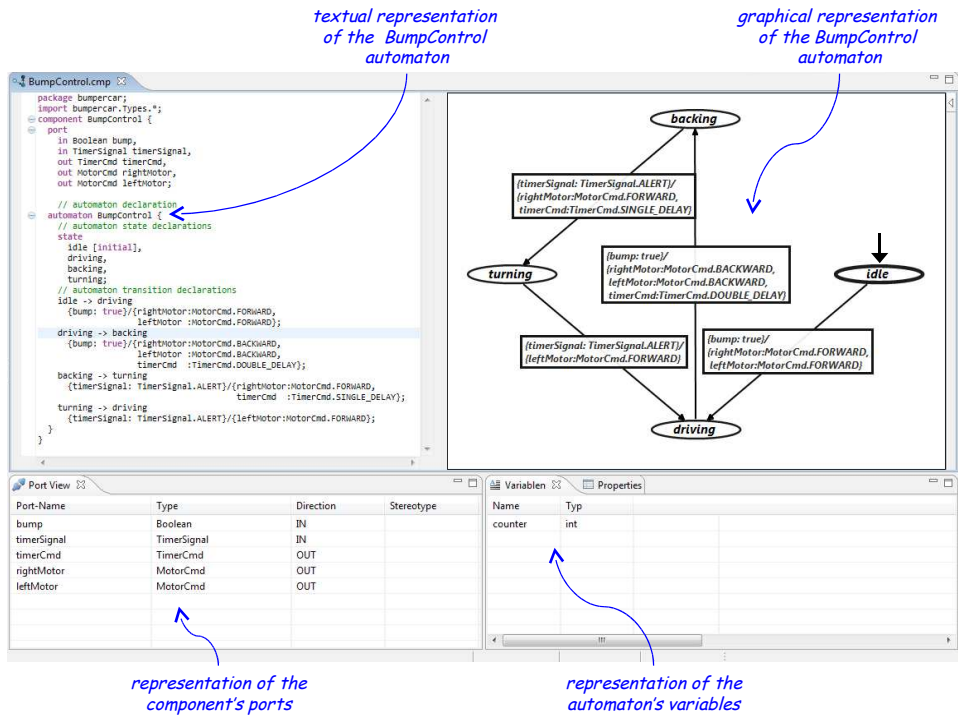

Figure 6: The combined textual and graphical editor for MontiArcAutomaton models featuring parallel textual and graphical editing of the same model.

For graphical editing, we have defined a generic automaton language using the EMF Ecore meta model. MontiArcAutomaton models are generated into models of a generic automaton language and displayed using the Eclipse Graphical Editing Framework[5]. We chose not to define a distinct I/O$^\omega$ automaton EMF language, as we want to embed other kinds of automata into MontiArc components in the future and thus would have to define new EMF Ecore languages for each kind.

---

[5]Eclipse Graphical Editing Framework: http://www.eclipse.org/gef/

## 4 MontiArcAutomaton: a Software Architecture Structure and Behavior Modeling Language

We use the ADL MontiArc [HRR12] to model CPS. MontiArc describes architectures using components and connectors. Components encapsulate subsets of the systems functionality and regulate control via explicitly defined interfaces, which in MontiArc are sets of directed typed ports. The type of a port determines the possible messages a component may receive or send on that port, and can be defined using UML/P class diagrams [Rum11]. Connectors effect and regulate the interaction of components.

MontiArc realizes the semantics of FOCUS [BS01] to describe component behavior. Components are treated as black-boxes that read input streams and produce output streams (message streams). These streams are the observable history of component interactions. A MontiArc component is either *atomic* and its behavior is defined explicitly or *composed* and its behavior is defined solely by the structural composition of the behaviors of its subcomponents. Models of composed components define the relations between subcomponents and connectors. MontiArc distinguishes between the definition of a component and its instantiation, supports powerful typing, instantiation, and parametrization mechanisms as described in [HRR12], but does itself not provide a language to model the behavior of its components explicitly. We thus have extended MontiArc by embedding I/O$^\omega$ [Rum96] as component definitions. The semantics of I/O$^\omega$ automata are stream processing functions. Our MontiCore implementation of this language is called MontiArcAutomaton.

### 4.1 The MontiArcAutomaton Language

We added local variables, states and transitions to MontiArc components to model component behavior. Variables can be used by automata to store and look up intermediate values. A transition connects a source state with a target state and has an optional guard, input block, and output block. Variables, states and transitions are only visible inside a component and all communication between components is made explicit via ports and connectors. I/O$^\omega$ automata do not feature hierarchical states because decomposition takes place on component level and thus reduces the complexity of component behavior description significantly. The input block of a transition defines patterns of messages and events received on incoming ports or stored in the local variables of the component that together with the guard activate the transition. A guard is a predicate over the messages on input ports and values stored in local variables. Guards in MontiArcAutomaton can be specified using OCL/P [Rum11, Sch12] a MontiCore implementation of OCL. The reaction of a component to an input is specified by the output block, which is a set of pairs of output ports and the (streams of) messages that are sent as a reaction to the input. It also may contain assignments to the local variables of a component.

Listing 7 displays an excerpt from the concrete syntax of the automaton of component `BumpControl` (see Figure 2). The state `idle` is an initial state and defines initial outputs

```
 1  component BumpControl {
 2    // Port declarations
 3    automaton {
 4      state
 5        idle [initial {rMotor:STOP,
 6                       lMotor:STOP}],
 7        driving, backing, turning;
 8
 9      idle -> driving {bump:true} / {rMotor:FORWARD,
10                                     lMotor:FORWARD};
11      // ... more transitions here
12  }}
```

Listing 7: An excerpt from the concrete syntax of the automaton of component `BumpControl`.

in ll. 5-6. The transition from state `idle` to state `driving` is shown in ll. 9-10. It only reads from one port but defines the output on multiple ports. In general, transitions may read from all incoming ports and send messages on all outgoing ports.

The semantics of MontiArcAutomaton is described as sets of stream processing functions (SPF) [Rum96, RR11]. In the case of a total and deterministic automaton the set is a singleton, otherwise each stream processing function (SPF) describes a different possible implementation of the desired system. The SPF corresponding to a MontiArc component maps one input stream bundle to one output stream bundle. The input stream bundle contains streams for each input port of the component and the output stream bundle contains streams for each output port of the component. For a formal definition of MontiArcAutomaton semantics see [Rum96] and the language report on the MontiArcAutomaton website [Mon12].

## 5 Code Generation from MontiArcAutomaton

The code generation framework of MontiCore uses FreeMarker templates and adds *template operators* and *template calculators* to transform the ASTs of models of MontiCore languages into implementations. Template operators provide the infrastructure for code generation, access the AST, call template calculators and include sub templates, e.g., for connectors or transitions. Templates consist of target language fragments and FreeMarker control structures. Template calculators perform complex operations and provide symbol table access that would complicate the templates or be impossible inside templates due to FreeMarker restrictions. The template operator also persists the results of calculations.

Listing 8 illustrates these concepts on the template for a single transition in Python. The template executes the method `getFrom()` on the current AST node (l. 1). Line 2 uses the FreeMarker directive `<#if>..</#if>` to determine and evaluate the transition's guard calling the template calculator `guardCalculator`. The result of this evaluation is stored in the field `guardExpression` of the `guardCalculator` and thus available in the template via the template operator. Afterwards, the template iterates over all input ports/channels (l. 5) and includes a template called `concatStream` for each input (l. 8).

```FreeMarker
if (self._state == ${op.getValue("enumName")}.${ast.getFrom()}
<#if op.callCalculator(guardCalculator)>
  and (${op.getValue("guardExpression")})
</#if>
<#foreach chin in ast.getChannel_in()>
  and (self._${chin.getInName()}.getCurrentValue() != None
      and self._${chin.getInName()}.getCurrentValue() ==
         ${op.includeTemplates(concatStream, chin.getInput())})
</#foreach>
  ):
<#foreach chout in ast.getChannel_out()>
  self._${chout.getOutName()}.setNextValue(
    ${op.includeTemplates(concatStream, chout.getOutput())})
</#foreach>
  self._state = ${op.getValue("stateEnumName")}.${ast.getTo()}
```

Listing 8: The template for the Python implementation of a single transition in FreeMarker.

Parts of the Python code generated from the automaton of component `BumperControl` (from Figure 2) are shown in Listing 9. The initial state and output of the component are set in ll. 4-6. The code of the transition from state `idle` to `driving` (shown in ll. 8-14) is generated based on the template from Listing 8.

```Python
class BumpControl(runtime.Component):

  def init(self):
    self._state = BumpControlState.idle
    self._rMotor.setCurrentValue(MotorCmd.STOP)
    self._lMotor.setCurrentValue(MotorCmd.STOP)

  def compute(self):
    if (self._state == BumpControlState.idle and
        (self._bump.getCurrentValue() != None and
         self._bump.getCurrentValue() == True) ):
      self._rMotor.setNextValue(MotorCmd.FORWARD)
      self._lMotor.setNextValue(MotorCmd.FORWARD)
      self._state = BumpControlState.driving
      # ... more transitions here
```

Listing 9: An excerpt from the Python class generated from the automaton inside component `BumpControl` (cf. Lst. 3 and 4)

Using this framework, we have developed four code generators by implementing new templates and reusing most of the MontiArcAutomaton dependent template calculators.

The next section explains how MontiCore supports reuse in code generator development.

### 5.1 Multiple Target Language Code Generation

MontiCore facilitates toolchain reusability as grammar, symbol table, and context conditions are part of the language front-end. They can easily be reused for every new code generator and editor. Figure 10 illustrates this reuse. Common workflows and code shared among the different back-ends (code generators) for MontiArcAutomaton, are packaged in the project `MontiArcAutomatonBECommons`. This way, code generation reuses calculators common to all back-ends (e.g., collection of states and transition) as well as symbol tables entries and context conditions generated and implemented for the different inherited and embedded modeling languages respectively.

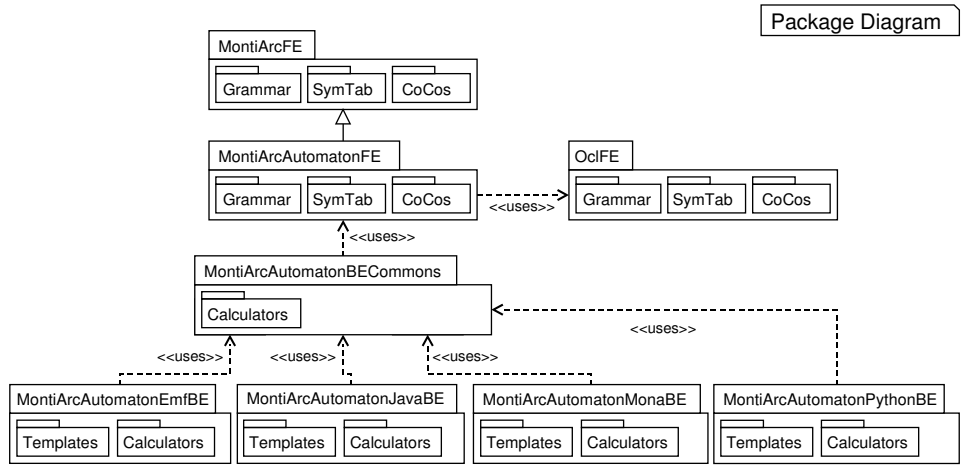

Figure 10: Dependencies between language front-ends (containing languages and context conditions) and back-ends (containing generators).

We have developed (code) generators from MontiArcAutomaton models to EMF Ecore (`MontiArcAutomatonEmfBE`), Java code (`MontiArcAutomatonJavaBE`), Mona (`MontiArcAutomatonMonaBE`) and Python (`MontiArcAutomatonPythonBE`). These generators reuse the parsers, symbol table generation, and context conditions of MontiArc, OCL/P, and MontiArcAutomaton. Overall, for these four code generators, we developed between one (EMF) and 3 (Mona) target language specific templates as well as between two (EMF) and five (Python) target language specific template calculators: as Python, for example, lacks explicit types, a new calculator for the parameter lists of method signatures had to be developed. The average template lengths – including comments – range from eleven lines (Mona) to $48$ lines (EMF). Table 11 quantifies this reuse. As all generators use similar structures (e.g., iteration over the automaton's transitions), we could identify and reuse seven common template calculators (provided by `MontiArcAutomatonBECommons`).

| Generator | Number of (additional) Calculators | Number of Templates | Average Template Length |
|---|---|---|---|
| MontiArcAutomatonBECommons | 7 | - | - |
| MontiArcAutomatonEMFBE | 1 | 2 | 48 |
| MontiArcAutomatonTSBE | 2 | 12 | 33 |
| MontiArcAutomatonMonaBE | 3 | 11 | 11 |
| MontiArcAutomatonPythonBE | 2 | 13 | 29 |

Figure 11: Key figures on calculator reuse in code generator development. The generator project `MontiArcAutomatonBECommons` contains generic MontiArcAutomaton calculators. The other projects add additional target platform specific calculators.

Also, the target languages pose some restrictions on valid models of the input language. For example, the code generation to Java supports generic port types not supported by the Mona code generation. As another example, the analysis using Mona heavily depends on underspecification mechanisms, e.g., undefined inputs and outputs and non-deterministic transitions. The Java and Python code generators do not translate these models properly. Thus, we have developed sets of context conditions that define language profiles of MontiArcAutomaton for different generators.

Besides toolchain reusability, MontiCore also facilitates language reusability as these may inherit from others (MontiArcAutomaton inherits from MontiArc) and embed other languages (MontiArcAutomaton embeds OCL/P). The new language can reuse context conditions, template calculators, and symbol tables from the inheriting or embedding language.

We want MontiArcAutomaton models to be platform independent: this no longer works, when dealing with hardware or API access. We therefore postulate the existence of (and have developed) a runtime library per target platform. This library contains component models that deal with target system interaction (e.g., hardware access, usage of other frameworks, etc.). For the robotics project, we implemented such a library consisting of components wrapping access to sensors and motors, such that the system engineers were able to model the systems without taking the underlying API into account.

### 5.2 Challenges: Equal Operational Semantics and Libraries

**Equal Operational Semantics** We have developed multiple code generators for the same language. During system development a single model is used with the EMF Ecore code generator for editing, the Mona code generator for validation and verification and the Java LeJOS as well as the Python ROS code generators for execution on different platforms. For the latter three translations it is of great importance that all models behave according to a single operational semantics (to the extend necessary to preserve properties established in prior verification and validation steps).

Enforcing a common operational semantics mutually affected the code generators. On one hand, the common operational semantics requires synchronization of communication and computation on distributed CPUs in our Java and Python implementations, since our implementation in Mona only supports synchronized components (the current translation can not model unbounded buffers). On the other hand, it also requires the modeling of null values on message streams in the Mona implementation. Furthermore, we had to choose a compatible encoding of messages and of communication protocols (partially implemented using hardware wrappers) between robots running LeJOS or ROS.

In the code generators presented here we have addressed the challenge of ensuring equal operational semantics by careful inspection of the code generators (facilitated by the common structure) and validation based on test cases.

**Libraries** Model and code libraries are important in every language to provide reuse based on common interfaces. The MontiCore symbol table framework [Vö11] implements a conform look-up and handling of components independently of their source (library or model) and implementation (generated or manual). All tools operating on the model level, e.g., the content assistants of our editors and context condition checks, thus fully support library components.

For our code generators pure model libraries are no problem since the complete implementation can be generated for any target language. As discussed in the previous section, we also employ library components that need target platform specific implementations we can not describe with the MontiArcAutomaton language. We never the less model these library components as MontiArcAutomaton models to provide consistent interfaces on the model level. In addition, we have implemented mechanisms to combine the generated code with manual implementations without modifying the generated code (e.g., the factory pattern and delegation). Since platform specific implementations, e.g., hardware access, are not contained in the models, this code has to be manually created for every target platform. The realization of components corresponding to hardware wrappers is still a challenging manual task in our Mona implementation.

## 6 Related Work

The AutoFOCUS [HST10, HF07] tool chain for model-based development of reactive, distributed systems supports modeling of logical architectures (similar to MontiArc), technical/platform architectures and deployment mappings. MontiArcAutomaton and AutoFOCUS share the same semantic domain FOCUS [BS01]. Behavior of AutoFOCUS components can be modeled using input/output automata similar to MontiArcAutomaton [HF07]. The AutoFOCUS tool chain contains code generators to multiple target platforms [HST10], e.g., to C code and to the theorem prover Isabelle/HOL. To the best of our knowledge the AutoFOCUS verification mechanisms do not support fully automated refinement checking as shown in [Kir11] and code generation to different robotics platforms as presented in this paper.

**Modeling Languages** The UML [OMG12] is a general purpose modeling language family consisting of 14 diagrams for structure, behavior and interaction modeling. Among these are *state machine diagrams*, to model the behavior of systems in terms of states and transitions, and *component diagrams*, to model the interaction of components via the ports of their interfaces. While these could be used to model architecture and behavior of CPS, the semantics of the embedding of state machines into components is not explicitly defined in UML.

SysML [FMS11] is a general-purpose modeling language based on UML. The language features several diagram types known from UML and introduces additional ones. Among these diagrams are UML *state machine diagrams* and *internal block diagrams* based on UML *composite structure diagrams*. While the latter, provided to model the internal structure of classes using blocks and ports, might be used to model a system architecture, the UML state machine diagrams still suffer the problems mentioned above.

**Language Workbenches** The MontiCore approach to language design is similar to the Eclipse Xtext [EB10] approach, which also uses EBNF-like grammars to generate parser, text editor eclipse plugin, and further tooling. The AST generated from Xtext parsers is EMF-based and code generation uses the Eclipse Xtend programming language[6] to define the templates used for code generation. Xtend augments Java with several concepts while compiling into interoperable Java source code. As Xtext requires Eclipse, the integration into tool chains is hampered. MontiCore offers Eclipse integration and in addition API and command line tools that are independent of the Eclipse framework.

The Meta Programming System (MPS)[7] follows a different approach, as it follows a projectional approach, i.e., lets the user directly edit a model's underlying AST. MPS uses a proprietary meta model and does neither provide Eclipse integration, nor means of automated toolchain integration. The code generation with MPS is implemented as a model-to-model transformation, which requires grammars of all target languages. Our approach does not require grammars of the target languages.

## 7 Conclusion

We have presented the MontiArcAutomaton modeling language to model architecture and behavior of CPS and illustrated how the development of CPS can be improved using modeling and verification techniques. The MontiArcAutomaton framework contains four code generators from platform independent MontiArcAutomaton models to Mona (formal analysis), EMF Ecore (graphical editing), and Java and Python (deployment).

Based on these code generators we have illustrated how the MontiCore DSL framework supports development and reuse of code generators, including (partial) reuse of symbol tables, context conditions, and template calculators.

Currently, we are working on a modeling language to specify the deployment of MontiArcAutomaton components to hardware. Furthermore, we are conducting an evaluation of the MontiArcAutomaton framework with master students.

---

[6]Xtend programming language: http://www.eclipse.org/xtend/
[7]Meta Programming System: http://www.jetbrains.com/mps/